\documentstyle[aps,multicol,prl,epsfig]{revtex}

\newcommand{\equ}[1]{(\protect\ref{#1})}
\newcommand{\jam}{\theta_\infty}
\newcommand{\si}{\sigma}
\newcommand{\pt}[1]{\frac{\partial {#1}}{\partial t}}
\newcommand{\G}[1]{G(#1, t)}
\newcommand{\D}{{\rm d}}

\newcommand{\Raya}{\vspace*{-0.25cm}\noindent\makebox[3.4truein]{\hrulefill}}
\newcommand{\Mycaption}[1]{\refstepcounter{figure}\protect\noindent%
  \protect\parbox{8.6cm}{\small FIG \thefigure. #1}}

\begin{document} 
\draft 

\title{Analytic model for the ballistic adsorption of polydisperse
  mixtures}

\author{Romualdo Pastor-Satorras\cite{newaddress}}
  
\address{Department of Earth, Atmospheric, and Planetary Sciences\\
Massachusetts Institute of Technology, Cambridge, Massachusetts 02139}

\date{\today}

\maketitle

\begin{abstract}
We study the ballistic adsorption of a polydisperse mixture of spheres
onto a line. Within a mean-field approximation, the problem can be
analytically solved by means of a kinetic equation for the gap
distribution. In the mean-field approach, the adsorbed substrate as
approximated as composed by {\em effective} particles with the {\em
same} size, equal to the average diameter of the spheres in the
original mixture. The analytic solution in the case of binary mixtures
agrees quantitatively with direct Monte Carlo simulations of the
model, and qualitatively with previous simulations of a related model
in $d=2$.
\end{abstract}

\pacs{PACS number(s): 68.45.Da, 81.15.-z, 82.70.Dd, 68.10.Jy}

\begin{multicols}{2}

\section{Introduction}

The adsorption of colloidal particles onto a surface is a subject of
considerable interest due to its many practical applications in fields
as diverse as physics, chemistry, biophysics, medicine, etc.
\cite{evans93}. Several models have been proposed so far, in an
attempt to understand the physical properties of the adsorbed phase.
In the random sequential adsorption (RSA) model
\cite{renyi58,feder80,schaaf88,senger91,ramsden93}, the adsorbing
particles are located at random positions on the surface.  If an
incoming particle overlaps with a previously adsorbed one, it is
rejected; otherwise, it becomes irreversibly adsorbed.  The RSA model
does not consider the transport of the particles, and focuses only on
the excluded volume effects.  It is thus a valid approximation when
the particles arrive at the surface purely by diffusion
\cite{senger91}.  In the ballistic model (BM)
\cite{meakin87,talbot92,thompson92}, when an incoming particle fails
to reach the surface directly, it is allowed to roll down over the
previously adsorbed ones, following the direction of the steepest
descent, until it reaches an equilibrium position. Particles that
eventually rest on the surface are irreversibly adsorbed; otherwise,
they are rejected.  The BM is therefore a good approximation to
describe adsorption in presence of strong interactions, attracting the
particles towards the surface \cite{senger93,schaaf95,pastor98a}.
 
In their original formulation, the aforementioned models, as well as
their main variations, consider essentially the adsorption of a {\em
  monodisperse} suspension, in which the adsorbing particles have all
the same size.  Real-life suspensions, however, always possess an
unavoidable degree of polydispersity.  For instance, in some
experimental situations the standard deviation of the particle size
distribution may be up to 5--10$\%$ of the mean particle size
\cite{wojta93,adamczyk97}. Under such conditions, the effects of
polydispersity may be indeed important.
  
The role of polydispersity has been studied in some detail in the RSA
model. Theoretical works have dealt with binary mixtures of particles
with greatly differing diameters \cite{talbot89}, power-law size
distributions \cite{krapivsky92,brilliantov96}, or general continuous
size distributions \cite{tarjus91}. Numerical simulations, on the
other hand, have been performed in a wider variety of conditions:
binary mixtures \cite{meakin92,bonnier92}, uniform size distributions
\cite{meakin92}, gaussian distributions \cite{adamczyk97,meakin92},
power-law distributions \cite{brilliantov96}, etc.

In the context of the BM, it is noteworthy the work of Senger {\em et
al.} \cite{senger93}, where the authors describe a Monte Carlo model
for the adsorption of a two-component mixture of hard spheres onto a
plane, where the particles are under the simultaneous influence of
diffusion and gravity.  This is indeed a mixed model, which reproduces
the standard RSA model in the limit of small particles, and the BM for
large particles. The results reported by Senger {\em et al.}, for
particles large enough to be well inside the BM regime, are
qualitatively similar to those found for the RSA of binary mixtures
\cite{meakin92}: The maximum fraction of surface covered by the
adsorbed particles---the {\em jamming limit} $\jam$---increases
monotonically with the concentration $p$ of large particles, with a
maximum in the limit $p\to1^-$ (i.e., $1-p$ arbitrarily small, but
non-zero). For $p=0$ and $p=1$, the coverage corresponding to
monodisperse adsorption is recovered.

In this paper we present an analytic model for the ballistic
adsorption of mixtures of spherical particles with different
diameters.  The model can be solved in a mean field approximation, by
studying the kinetics of the gap density function \cite{talbot92}.
Within this approach, we are able to derive a generic equation for an
effective gap distribution. To test our equation, we solve it
explicitly in the simplest case of a binary mixture. The analytic
results obtained for the density at jamming $\jam$ match the findings
of direct Monte Carlo simulations of the model. Moreover, the
qualitative behavior of $\jam$ predicted by our model is the same as
the reported by Senger {\em et al.}  \cite{senger93}.

\section{Model}

Our model considers the adsorption onto a line of a polydisperse
mixture whose degree of polydispersity is characterized, in general,
by a continuous distribution of sizes $\rho(\si)$.  The quantity
$\rho(\si)\D \si$ is defined as the fraction (bulk concentration in
the infinite reservoir from which the particles are drawn) of
spheres with diameter between $\si$ and $\si+\D \si$. We assume $\rho$
to be normalized to 1. Thus, for a monocomponent solution of particles
of size $\si_o$, we have $\rho(\si) = \delta(\si - \si_o)$. The
particles arrive at the line at rates $k(\si)$ per unit length per
unit time. Assuming that the adsorbed substrate interacts with the
incoming particles only through excluded volume effects, we can select
the appropriate units of time and set $k(\si)\equiv\rho(\si)$.  Under
these conditions, the problem translates into the sequential
adsorption of particles of size $\si$, selected with a probability
density $\rho(\si)$.

\begin{figure}[t]
  \centerline{\epsfig{file=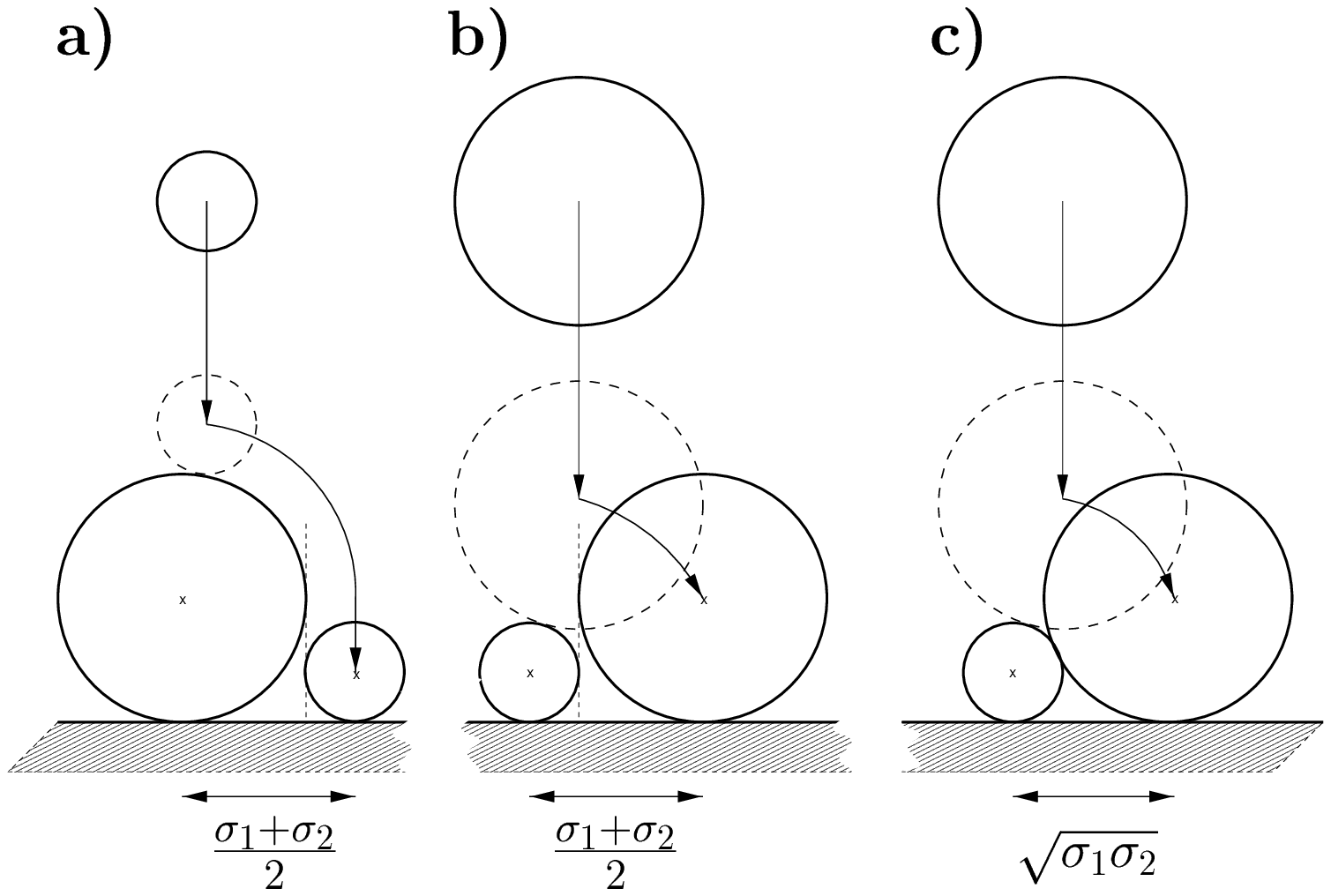, width=8cm}}
  \Mycaption{Landing configurations for particles of different sizes
    $\si_1$ and $\si_2>\si_1$.}
  \label{configurations}
\end{figure}

When an incoming particle lands over a preadsorbed one of exactly the
same size, the adsorption rules are identical to the standard BM
\cite{talbot92}.  Figure~\ref{configurations} depicts the possible
configurations involving particles of different diameters $\si_1$ and
$\si_2$, with $\si_1 < \si_2$. When a small particle rolls over a
large one, the former finally falls on the surface and, after it is
adsorbed, the centers of both particles are separated a horizontal
distance $\Delta = (\si_1+\si_2)/2$; Fig.~\ref{configurations}(a)
represents this case.  When a large particle rolls over a small one,
the rule adopted in our model is the one represented in
Fig.~\ref{configurations}(b), in which, after rolling, the centers of
the two particles are also separated a distance $\Delta$.

The adoption of the rule pictured in Fig.~\ref{configurations}(b)
represents a major simplification in the model. It could be possible
to argue that, in a more realistic treatment, the final configuration
involving a large particle rolling over a small one should be the one
depicted in Fig.~\ref{configurations}(c). The surfaces of the
particles are tangent after adsorption in this case, and their centers
are separated a horizontal distance $\Delta_R=\sqrt{\si_1 \si_2}$.
Both rules can be easily implemented in a numerical simulation.
However, the prescription \ref{configurations}(c) imposes an essential
asymmetry among particles of different size. First of all, in our
model, as defined by rules \ref{configurations}(a) and
\ref{configurations}(b), the final result of an adsorption event
involving two spheres of different diameter is independent of the
order in which the particles reach the surface. As a consequence, our
model does not allow for ``overhangs''; this means that, if $n(\si)$
is the density of adsorbed particles of size $\si$, then the fraction
of covered surface $\theta$ is simply given by $\theta=\int \D \si \si
n(\si)$. This simple expression obviously does not hold in a model
defined with the rule \ref{configurations}(c).

These ``abelian'' properties are eventually responsible for our model
being analytically tractable.

\section{General Mean-Field Equation}

The model defined in the previous section can be analyzed by studying
the density function of gaps---holes between two consecutive adsorbed
particles.  Let us define $\G{x} \D x$ as the number of gaps with a
length between $x$ and $x+\D x$ present at time $t$, per unit length
of substrate.  The time evolution of $G$ is obtained as a balance
equation for the creation and destruction of gaps caused by a single
adsorption event \cite{talbot92}.  Given $G$, the fraction of covered
surface is defined by
\begin{equation}
  \label{eq:density}
  \theta(t) = 1 - \int_0^\infty x \G{x} \D x,
\end{equation}
and, from here, we obtain the jamming limit as $\jam =
\lim_{t\to\infty} \theta(t)$.

In the case of the ballistic adsorption of a monodisperse solution of
spheres of diameter $\si_o$, the equations for the density of gaps are
\cite{talbot92}

\end{multicols}

\widetext
\Raya\hfill

\begin{eqnarray}
  \pt{\G{x}} & = & -(x+\si_o) \G{x} + 2 \si_o \G{x+\si_o} + 2
  \int_{x+\si_o}^\infty \G{y} \D y, \qquad {\rm for} \quad x>\si_o;
  \label{eq:single1} \\
  \pt{\G{x}} & = & 2 \si_o \G{x+\si_o} + 2
  \int_{x+\si_o}^\infty \G{y} \D y, \qquad  {\rm for} \quad
  x<\si_o.  \label{eq:single2}
\end{eqnarray}
The solution of \equ{eq:single1}--\equ{eq:single2} is
\begin{eqnarray*}
  \G{x} & = & e^{-(x+\si_o)t} t^2 F(\si_o t) \exp\left\{2 (1-e^{-\si_o
      t})\right\}, \qquad {\rm for} \quad x>\si_o; \\
  \G{x} & = & 2 \int_0^t \D u \;u(1+\si_o u) e^{-(x+2\si_o)u} F(\si_o u)
  \exp\left\{ 2 (1-e^{-\si_o u})\right\}, \qquad  {\rm for} \quad
  x<\si_o,
\end{eqnarray*}
\hfill\Raya
\begin{multicols}{2}
where we have defined the auxiliary function
\begin{equation}
  \label{eq:definitionF}
  F(t)=\exp\left\{ -2 \int_0^t \frac{1-e^{-z}}{z} \D z \right\}.
\end{equation}

For a polydisperse mixture, the naive application of this approach
becomes considerably more involved. After a moment's reflection, it is
easy to realize that, in this case, the final configuration resulting
from an adsorption event taking place on a given gap, depends on the
sizes of the particles defining the boundaries of that gap.  We should
accordingly deal with a continuous set of functions $G_{\si',\si''}(x,
t)$, defined as the densities of gaps created between particles of
size $\si'$ and $\si''$, for $\si', \si'' \in [0, \infty]$ . An
enumeration of all the possible events occurring when adsorbing
spheres of size $\si$ at rate $\rho(\si)$, would lead to a system of
exact coupled integro-differential equations for the magnitudes
$G_{\si',\si''}$, that would completely determine the dynamics of the
process. The magnitude of this task, especially when dealing with
continuous size distributions $\rho(\si)$, seems to preclude any
chance for an exact solution.

Fortunately, however, a great deal of insight can be gained by seeking
for a {\em mean-field} type of solution, based in the following
argument: When the particles are free in the suspension, they are {\em
distinguishable} and interact differently with the adsorbed phase,
depending on their size. However, once they have been adsorbed, we can
assume that they become {\em indistinguishable}, in the sense that the
adsorbed particles interact with the incoming particles as if the
former were all {\em equal}, with the same average diameter $\bar{\si}
= \int \si \rho(\si) \D \si$.  In other words, we can approximate the
adsorbed phase as composed by {\em effective} particles of the same
size $\bar{\si}$, interacting with incoming particles of size $\si$.
Assuming this simplification, we need only a single {\em effective gap
distribution} $G$, defined by the gaps bounded by the adsorbed
effective particles.

We remark the important fact that the aforementioned mean-field
approximation {\em does not} imply at all that the density of adsorbed
particles is proportional to the bulk density, $n(\si)
\propto \rho(\si)$. This relation, which can be true at the first
stages of the adsorption process, does not hold close to the
jammed---saturated---state. This last statement is most easily seen
in binary mixtures (see Sec.~\ref{binary}).

The kinetic equation for the effective gap density  can be written
in the generic form

\end{multicols}

\widetext
\Raya\hfill
\begin{equation}
  \pt{\G{x}} = - \int_0^x \D \si \, \rho(\si) (x+\bar{\si}) \G{x} + 2
  \int_0^\infty \D \si \, \rho(\si) \left(\frac{\si}{2} +
  \frac{\bar{\si}}{2} \right) \G{x+\si} 
  + 2 \int_0^\infty \D \si \,
  \rho(\si) \int_{x+\si}^\infty \D y \, \G{y}.  
  \label{eq:generic}
\end{equation}

\hfill\Raya
\begin{multicols}{2}

The origin of the different terms in \equ{eq:generic} is the
following: The destruction of gaps of length $x$ is due to the landing
of a particle of size $\si$ on any point of an interval of length
$x+\bar{\si}$ centered on the gap. After averaging over the
distribution of incoming particles of size $\si<x$, we obtain the
first term in \equ{eq:generic}.  A gap of length $x$ can be created by
the impact of particles of size $\si$ on either of the particles of
effective size $\bar{\si}$ defining a gap of length $x+\si$.  These
events, which happen at rate $\rho(\si)$, account for the second term
in \equ{eq:generic}.  The last term is due to the averaged creation of
gaps of length $x$ by direct deposition of a particle of size $\si$
onto a gap of length $y>x+\si$.  We remark again that
Eq.~\equ{eq:generic} owes its relatively simple form to the choice of
the ``abelian'' rule \ref{configurations}(b) in the definition of the
model. A much more complex expression would have been obtained with
rule \ref{configurations}(c).

Equation~\equ{eq:generic} can be expressed in a more compact way by
integrating by parts its last term. Defining the distribution function
$\Psi(x)= \int_0^x \rho(\si) \D \si$, we obtain 
\begin{eqnarray}
  \lefteqn{\pt{\G{x}} = - \Psi(x)(x+\bar{\si}) \G{x}} \nonumber \\
  &+& \int_0^\infty \D \si \,
  \Big[ \rho(\si) (\si + \bar{\si}) + 2 \Psi(\si) \Big] \G{x+\si}.
  \label{eq:compact}
\end{eqnarray}

Equation~\equ{eq:compact} is the final expression of the mean-field
theory for our model of polysdisperse ballistic adsorption.  As a
consistency check, we consider the trivial scenario of a monodisperse
suspension. In this case, by setting $\rho(\si)=\delta(\si-\si_o)$ and
$\Psi(x)=\Theta(x-\si_o)$, where $\Theta$ is the Heaviside step
function, we immediately recover the equations for a single-size
distribution, as given by Eqs.~\equ{eq:single1} and \equ{eq:single2}.

\section{Binary mixtures}
\label{binary}

In order to test the validity of our mean-field theory, we proceed now
to solve explicitly Eq.~\equ{eq:compact} in the case of a binary
mixture, composed by particles of size $\si_1=1$, which adsorb onto
the surface at rate $\phi_1$, and particles of size $\si_2=r>1$,
adsorbing at rate $\phi_r=1-\phi_1$.  As an aside, in this simple
setting we can estimate the variations in the jamming limit due to the
adoption of rule \ref{configurations}(b) instead of
\ref{configurations}(c).  One can expect that, for small values of
$r$, the outcome of both models should be similar. Indeed, numerical
simulations show that, for values of $r<2$, the difference between
prescriptions is always less than $1\%$, for all values of $\phi_r$.

The density function for a binary mixture has the form
$\rho(\si)=\phi_1\delta(\si-1) + \phi_r\delta(\si-r)$, whereas the
distribution function is $\Psi(x)= \phi_1 \Theta(x-1)\Theta(r-x) +
\Theta(x-r)$, and the average size $\bar{\si}=\phi_1 + r \phi_r$. By
inserting these expressions into \equ{eq:generic} or \equ{eq:compact},
we obtain the following set of equations:

\end{multicols}

\widetext
\Raya\hfill

\begin{eqnarray}
  \pt{\G{x}} & = & -(x+\bar{\si}) \G{x} + \phi_1(1+\bar{\si}) \G{x+1}
  + \phi_r(r+\bar{\si}) \G{x+r}  \nonumber \\
  &+& 2 \phi_1 \int_{x+1}^\infty \G{y}
  \, \D y +  2 \phi_r \int_{x+r}^\infty \G{y} \, \D y,  
  \qquad {\rm for} \quad x>r; \label{eq:large}\\
  \pt{\G{x}} & = & -\phi_1 (x+\bar{\si}) \G{x} + \phi_1(1+\bar{\si})
  \G{x+1} 
  + \phi_r(r+\bar{\si}) \G{x+r}  \nonumber \\
  &+& 2 \phi_1 \int_{x+1}^\infty \G{y}
  \, \D y +  2 \phi_r \int_{x+r}^\infty \G{y} \, \D y,
  \qquad {\rm for} \quad 1<x<r;  \label{eq:intermediate}\\
  \pt{\G{x}} & = & \phi_1(1+\bar{\si})
  \G{x+1} 
  + \phi_r(r+\bar{\si}) \G{x+r}  \nonumber \\
  &+& 2 \phi_1 \int_{x+1}^\infty \G{y}
  \, \D y +  2 \phi_r \int_{x+r}^\infty \G{y} \, \D y,
  \qquad {\rm for} \quad  0<x<1. \label{eq:small}
\end{eqnarray}

We observe that, for a binary mixture, one could in principle try to
solve {\em exactly} the model, by determining the rate equations for
the densities of gaps delimited by particles of size $1$ and $r$,
namely $G_{1,1}$, $G_{r,r}$, and $G_{1,r}$. However, in this case one
would end up with a set of nine coupled equations. The simplification
achieved through the mean-field theory is here evident.

We consider in particular the case $1<r<2$. To solve the kinetic
equations, we seek in \equ{eq:large} a solution of the form $\G{x} =
e^{-(x+\bar{\si})t} H(t)$. With this substitution, we are led to the
equation for $H(t)$:
\begin{equation}
  \frac{\D \ln H}{\D t} =  \phi_1  \left[ (1+\bar{\si})  +
    \frac{2}{t} \right] e^{-t} + \phi_r \left[ (r+\bar{\si})  + 
    \frac{2}{t} \right] e^{-rt} .
  \label{eq:partialeq}
\end{equation}
The solution of \equ{eq:partialeq}, with the initial condition
$H(0)=0$, is
\begin{equation}
  H(t)=t^2 \exp \big\{  \phi_1 (1+\bar{\si})(1-e^{-t}) \big\} 
   \exp \big\{\phi_r (r+\bar{\si})(1-e^{-rt})/r \big\}  
  [F(t)]^{\phi_1} [F(rt)]^{\phi_r},
  \label{eq:solutionh}
\end{equation}
where $F(t)$ is defined in \equ{eq:definitionF}. Upon
substituting this result into \equ{eq:intermediate}, we look for a
solution of this equation of the form $\G{x} = e^{-\phi_1
(x+\bar{\si})t} Q(x,t)$. The equation determining $Q$ is
\begin{equation}
  \pt{Q(x,t)}= e^{-\phi_r (x+\bar{\si})t} \frac{\D H(t)}{\D t},
\end{equation}
from which $Q(x,t)$ is obtained by direct integration, together with
the initial condition $Q(x,0)=0$: 
\begin{equation}
  Q(x,t)= e^{-\phi_r (x+\bar{\si})t} H(t) + \phi_{r} (x+\bar{\si})
  \int_{0}^{t} \D u \, e^{-\phi_r (x+\bar{\si})u} H(u).
\end{equation}
Finally, by substituting the solutions of \equ{eq:large} and
\equ{eq:intermediate} into the appropriate range of values of $x$ in
Eq.~\equ{eq:small} (and taking into account that $r<2$), we can
directly integrate this equation. Using Eq.~\equ{eq:density}, and
after performing some algebraic manipulations, we obtain the density
of adsorbed particles as a function of time:
\begin{eqnarray*}
  \theta(t) &=& \int_0^t \D u \, H(u) {\cal F}_1 (u) +
  \phi_r \int_0^t \D u \, H(u) {\cal F}_2 (\phi_1 t + \phi_r u)\\
  &+&  \phi_1 \phi_r (1+\bar{\si}) \int_0^t \D u \, H(u) 
  \int_u^t \D v \, {\cal F}_3  (\phi_1 v + \phi_r u) 
  +
  2 \phi_1 \phi_r \int_0^t \D u \, H(u) 
  \int_u^t \D v \, {\cal F}_4 (\phi_1 v + \phi_r u),
\end{eqnarray*}
where we have introduced the auxiliary functions 
\begin{eqnarray*}
  {\cal F}_1 (z) &=& \frac{e^{-\bar{\si} z}}{z^3} \left\{ \left[
      2\phi_r  +(1+\phi_r + \phi_r \bar{\si}) z +  (\bar{\si}+1) z^2
    \right] e^{-z}  - \left[ 2\phi_r + \phi_r (r+\bar{\si}) z
    \right] e^{-rz} \right\}, \\
  {\cal F}_2 (z) &=& \frac{e^{-\bar{\si} z}}{z^3} \left\{ - \left[
      2 + (\bar{\si}+2) z + (\bar{\si}+1) z^2 \right] e^{-z} + \left[ 
      2 + (\bar{\si}+2r)z +r(\bar{\si}+r) z^2 \right] e^{-rz}
  \right\}, \\
  {\cal F}_3 (z) &=& \frac{e^{-\bar{\si} z}}{z^3} \left\{ - \left[ 
      2 + (\bar{\si}+1) z  \right] e^{-z} + \left[ 2 +
      (\bar{\si}+2r-1) z + (r-1)(r+\bar{\si})z^2 \right] e^{-rz}
  \right\}, \\
  {\cal F}_4 (z) &=& \frac{e^{-\bar{\si} z}}{z^4} \left\{ - \left[ 3 +
(\bar{\si}+1) z \right] e^{-z} \right. \\
&+& \left. \left[ 3 + (\bar{\si} + 3 r - 2) z +
\frac{1}{2}(r-1)(3r + 2 \bar{\si}-1) z^2 +
\frac{1}{2}(r-1)^2(r+\bar{\si})z^3 \right] e^{-rz} \right\}.
\end{eqnarray*}
\hfill\Raya

\begin{multicols}{2}

\begin{figure}[t]
  \centerline{\epsfig{file=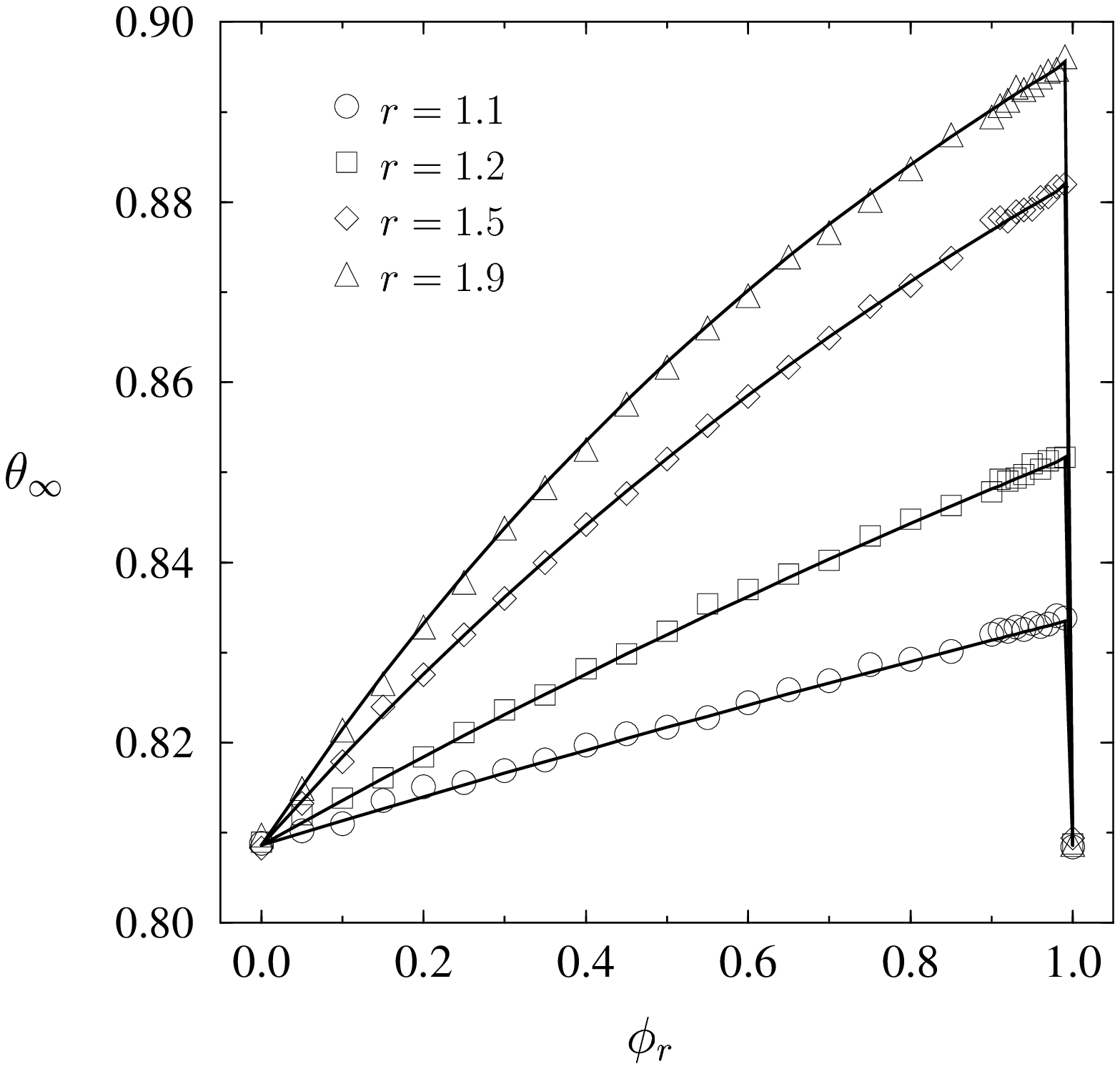, width=8cm}}
  \Mycaption{Jamming limit as a function of the concentration fraction
    $\phi_r$ of large particles, for different values of the diameter
    ratio $r$. Comparison between numerical simulations (hollow
    symbols) and the mean-field prediction (full lines).}
  \label{theory}
\end{figure}

We can estimate the theoretical predictions of this mean-field
solution by numerically integrating the previous expression in the
limit $t\to\infty$.  Figure~\ref{theory} shows in full lines the
results of the integration for different values of $r$. The symbols
represent data obtained from direct Monte Carlo simulations of the
model on a line of length $1000$ with periodic boundary conditions. We
observe that the predictions of the mean-field theory are in excellent
agreement with the numerical simulations.

From Fig.~\ref{theory} we conclude that, for $\phi_r<1$, the jamming
limit is an increasing monotonic function of this variable.  For
$\phi_r=0$ or $\phi_r=1$ (only small or large particles,
respectively), we recover, for any $r$, the prediction of the standard
BM model, $\jam^{BM}\simeq0.808$ \cite{talbot92}.  For small values of
$\phi_r$, $\jam$ grows linearly, $\jam\simeq\jam^{BM} + \alpha(r)
\phi_r$, with an slope $\alpha(r)$ that increases with $r$. The value
of the slope at the origin can be easily estimated by Taylor expanding
the expression for $\theta(t)$. The jamming limit exhibits a maximum
located at $\phi_r\to1^{-}$, in qualitative agreement with the
findings of Senger {\em et al.} \cite{senger93}. The actual value of
the maximum $\jam^{max}(r)$ is an increasing function of $r$, with an
apparent tendency to saturate at large $r$. In the limit $\phi_r\to1$,
and for $r\gg1$, we can estimate the limiting value of $\jam^{max}(r)$
\cite{meakin92}: In this limit, the large particles cover first a
fraction of surface $\jam^{BM}$ of the line, leaving free a surface
$1-\jam^{BM}$ that is afterwards covered until jamming by the small
particles. The total coverage is therefore bounded by
$\jam^{max}(r)\leq\jam^{BM} +
(1-\jam^{BM})\jam^{BM}=\jam^{BM}(2-\jam^{BM})\simeq0.96339$. Monte
Carlo simulations confirm this extreme, yielding the value
$\jam^{max}=0.964\pm0.001$ for $r=20$ and $\phi_r=0.99$.

\section{Conclusions}

To sum up, in this paper we have presented an extension of the
classical ballistic model \cite{meakin87,talbot92,thompson92},
describing the ballistic adsorption onto a line of a polydisperse
mixture of spherical particles of different sizes $\si$, present with
a bulk concentration $\rho(\si)$.  The model is solved by means of a
mean-field-like equation for the gap density, which approximates the
adsorbed phase as composed by effective particles, all with the same
average diameter $\bar{\si} = \int \D \si \, \si \rho(\si)$,
interacting with incoming particles of variable size. To check our
mean-field approximation, we have explicitly solved the case of a
binary mixture.  The perfect match of the theoretical solution with
the numerical simulations confirms the validity of the mean-field
approximation, at least for this particular case.  Our findings agree
also with numerical simulations of a related model in two dimension
\cite{senger93}. On theoretical grounds, the proposed mean-field
approach could be a first step to deal with more complex situations,
as, for example, in higher dimensionalities, where the assumption of
an effective layer of adsorbed particles would be more reasonable, or
in the case of adsorption onto a substrate initially covered with
impurities.

\acknowledgments

This work has been financially supported by a grant from the
Ministerio de Educaci\'{o}n y Cultura (Spain). I would like to thank
Prof. M. Rub\'{\i} for helpful discussions and Dr. M.C. Miguel for a
careful reading of the manuscript.

\end{multicols}

\end{document}